\documentclass[reprint,amsmath,superscriptaddress,amssymb,aps,pra,twocolumn,showpacs,showkeys]{revtex4-1}
\usepackage{graphicx}
\usepackage{enumitem}
\usepackage{ifthen}
\usepackage{boolexpr}

\begin{document}

\author{N.\,A.~Borshchevskaia}\email{borschxyz@gmail.com}
\affiliation{Faculty of Physics,  M.~V.~Lomonosov Moscow State University, Moscow, Russia}%
\author{K.\,G.~Katamadze}
\affiliation{Faculty of Physics,  M.~V.~Lomonosov Moscow State University, Moscow, Russia}
\affiliation{Institute of Physics and Technology, Russian Academy of Sciences, Moscow, Nakhimovsky prospect, 34, Russia}
\affiliation{National Research Nuclear University MEPhI, Moscow, Kashirskoe shosse, 31, Russia}
\author{S.\,P.~Kulik}
\affiliation{Faculty of Physics,  M.~V.~Lomonosov Moscow State University, Moscow, Russia}
\author{S.\,N.~Klyamkin}
\affiliation{Chemistry Department, M.~V.~Lomonosov  Moscow State University, Moscow, Russia}
\author{S.V.Chuvikov}
\affiliation{Chemistry Department, M.~V.~Lomonosov  Moscow State University, Moscow, Russia}
\author{A.\,A.~Sysolyatin}
\affiliation{A.~M.~Prokhorov General Physics Institute, Russian Academy of Sciences, Moscow, Russia}
\author{S.\,V.~Tsvetkov}
\affiliation{A.~M.~Prokhorov General Physics Institute, Russian Academy of Sciences, Moscow, Russia}
\author{M.\,V.~Fedorov}
\affiliation{A.~M.~Prokhorov General Physics Institute, Russian Academy of Sciences, Moscow, Russia}

\date{\today}
\title{Luminescence in germania-silica fibers in 1-2~$\boldsymbol{\mu}$m region}

%
%
%




\begin{abstract}
We analyze the origins of the luminescence in germania-silica fibers with high germanium concentration (about 30~mol.\%
$\rm  GeO_2$) in the region 1-2~$\boldsymbol{\mu}$m with a laser pump at the wavelength 532~nm. We show that such fibers demonstrate the high level of luminescence which  unlikely allows the observation of photon triplets, generated in a third-order spontaneous parametric down-conversion process in such fibers. The only efficient approach to the luminescence reduction is the hydrogen saturation of fiber samples, however, even in this case the level of residual luminescence is still too high for three-photon registration.
\end{abstract}


\maketitle
%

\section{Introduction}
The discovery of an effective source of correlated photon pairs 
based on the spontaneous parametric down-conversion \cite{Akhmanov1967,Harris1967}
gave an 
 impetus to the growth of quantum optics and quantum information. Thus the development an effective three-photon source can become the next big step in this progress. Generation of GHZ-states \cite{Zeilinger1992}, study the features of three-photon polarization states \cite{
Fedorov2015}, heralded generation of entangled biphotons are possible applications of such three-photon source.

The natural way of obtaining three-photon states is the similar process, third-order spontaneous parametric down-conversion (TOSPDC),
 which can be explained phenomenologically  as a spontaneous decay of a laser pump photon into photon triplets (threephotons) due to the cubic susceptibility. A lot of attempts of threephoton generation were made recently, however, until now they were obtained only by cascaded processes at second-order susceptibility \cite{Hamel2014, Hubel2010} by the addition of third photon \cite{Rarity1999, Mikami2005} or by postselection \cite{Bouwmeester1999, Eibl2004}.
Fibers are considered to be the most suitable nonlinear media for the direct triplet generation  due to its potentially large interaction length, limited mode structure and possibility of mode dispersion characteristics control during the design process: tapered \cite{Corona2011}, microstructural \cite{Cavanna2016}, conventional highly doped germania-silica \cite{Richard2011, Borne2015}.

For the latter fiber type S. Richard et al. \cite{Richard2011} estimated the triplet generation efficiency in spontaneous scattering for 1~W pump at the wavelength 532~nm in the 1~m length fiber to be as high as 0,2 triplet/s which is quite realistic to obtain in the experiment. Besides, A.~Borne et al. \cite{Borne2015} obtained the third harmonic at 516~nm in germanuim-doped fibers, which is the opposite process with similar phase-matching conditions but a much higher efficiency.
Although the third harmonic was also registered in our germania-silica fibers (see Sec.\ref{Exp})
in this letter we demonstrate that triplets, generated in spontaneous process in such fibers, can't be registered because of the bright luminescence with  several orders higher intensity than the expected triplet generation rate.
\begin{figure}[h]
\centering
\fbox{\includegraphics[width=\linewidth]
{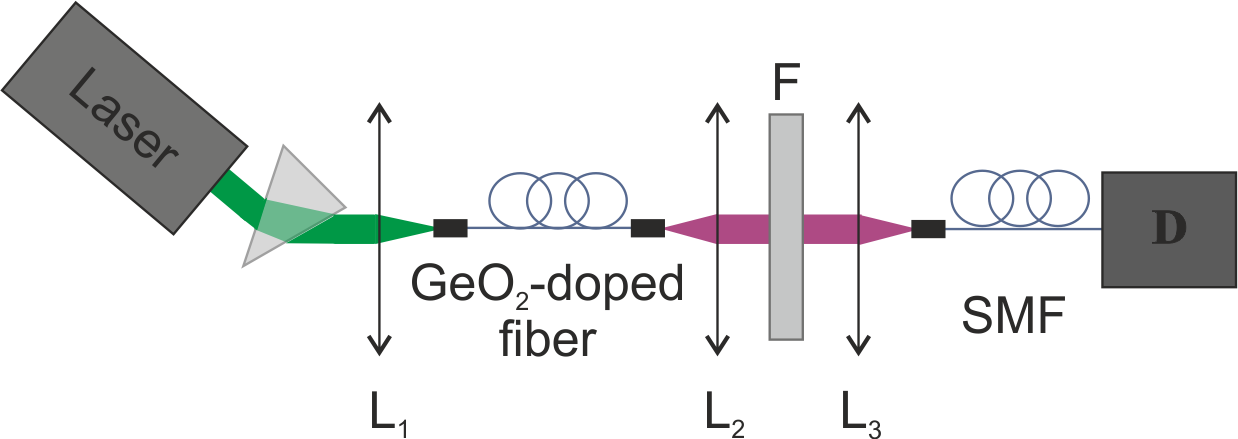}}
\caption{Experimental setup for luminescence registration.}
\label{fig:setup}
\end{figure}
At the same time, the parasitic luminescence is broadband, as well as triplets which are expected to be observed in the nondegenerate regime, moreover their spectral regions are similar.
The reason is that even if during the drawing process the fiber parameters are set to provide the degenerate phase-matching the minimal fluctuation of the fiber diameter at the level of 0,5\%
lead to nondegenerated threephoton generation in the region of the order 100~THz (1~$\mu$m) around the degenerate wavelength.

This letter is devoted to the study of such luminescence as well as possible ways of its elimination.

\section{Experiment}\label{Exp}
The experimental setup for measuring the luminescence rate is illustrated in Fig.~\ref{fig:setup}.
40~mW-diode laser at 532~nm was served as a pump. After passing through germanium-doped fiber the pump was eliminated by the filter \textit{F} which transmitted radiation in the region 0,6-2~mkm. The luminescence was registered by the counting \mbox{InGaAs} detector IDQuantique-210 with quantum efficiency 2,5\%,
its average pulse count rate was proportional to the average signal intensity. The light was focused into the fiber samples and then coupled with the detectors by lenses $L_1$, $L_2$ and $L_3$.

\begin{figure}[hp]
\centering
\fbox{\includegraphics[width=\linewidth]
{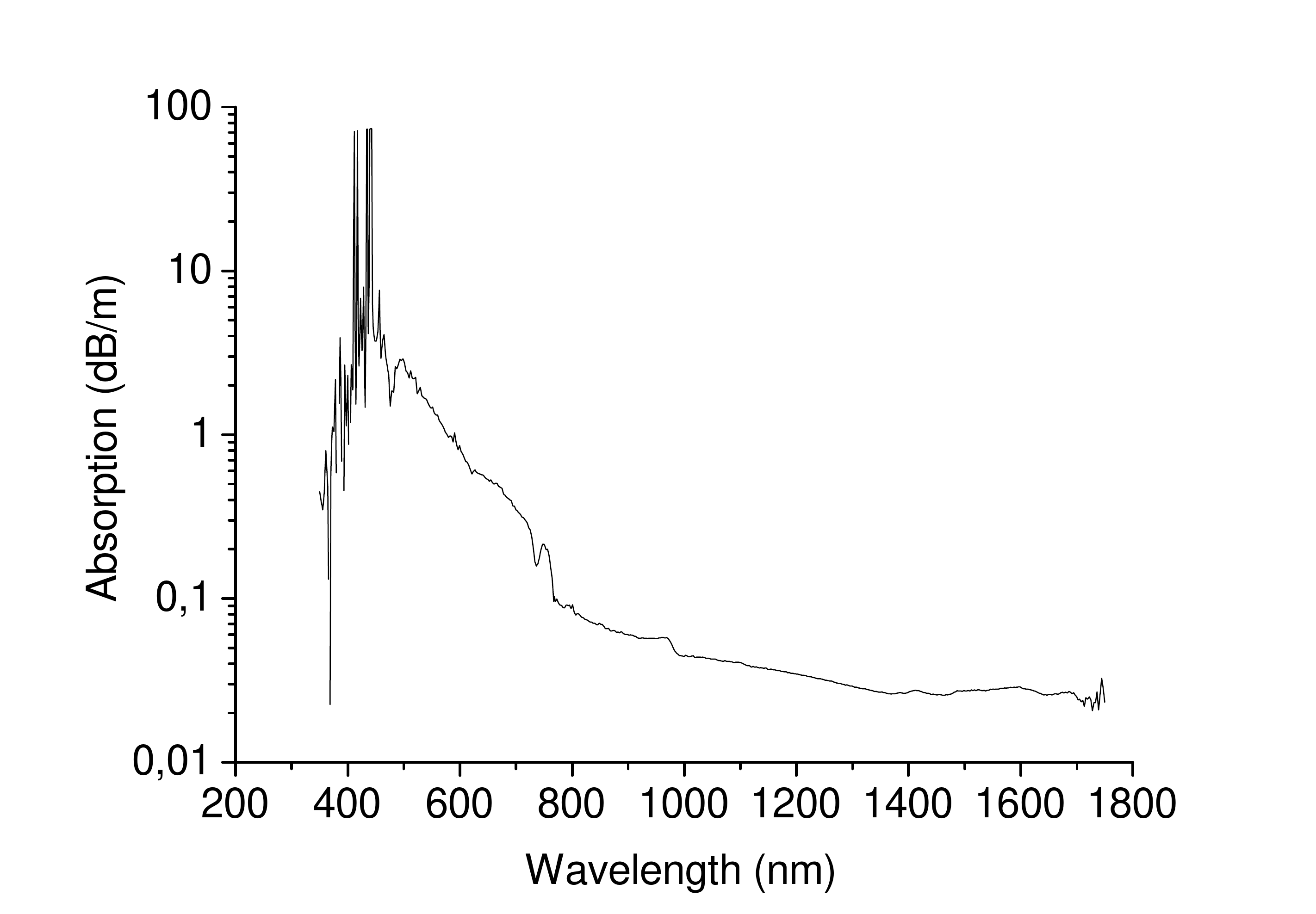}}
\caption{Absorption in fiber samples used in experiments.}
\label{fig:Losses}
\end{figure}

We studied germania-silica fibers highly doped with $\rm GeO_2$ (31~mol.\% $\rm GeO_2$),
the same as in \cite{Tsvetkov2016}. They absorb in UV and visible region with maximum at 400-500~nm (Fig.~\ref{fig:Losses}).

\begin{figure}[!bp]
\centering
\fbox{\includegraphics
[scale=0.28]
{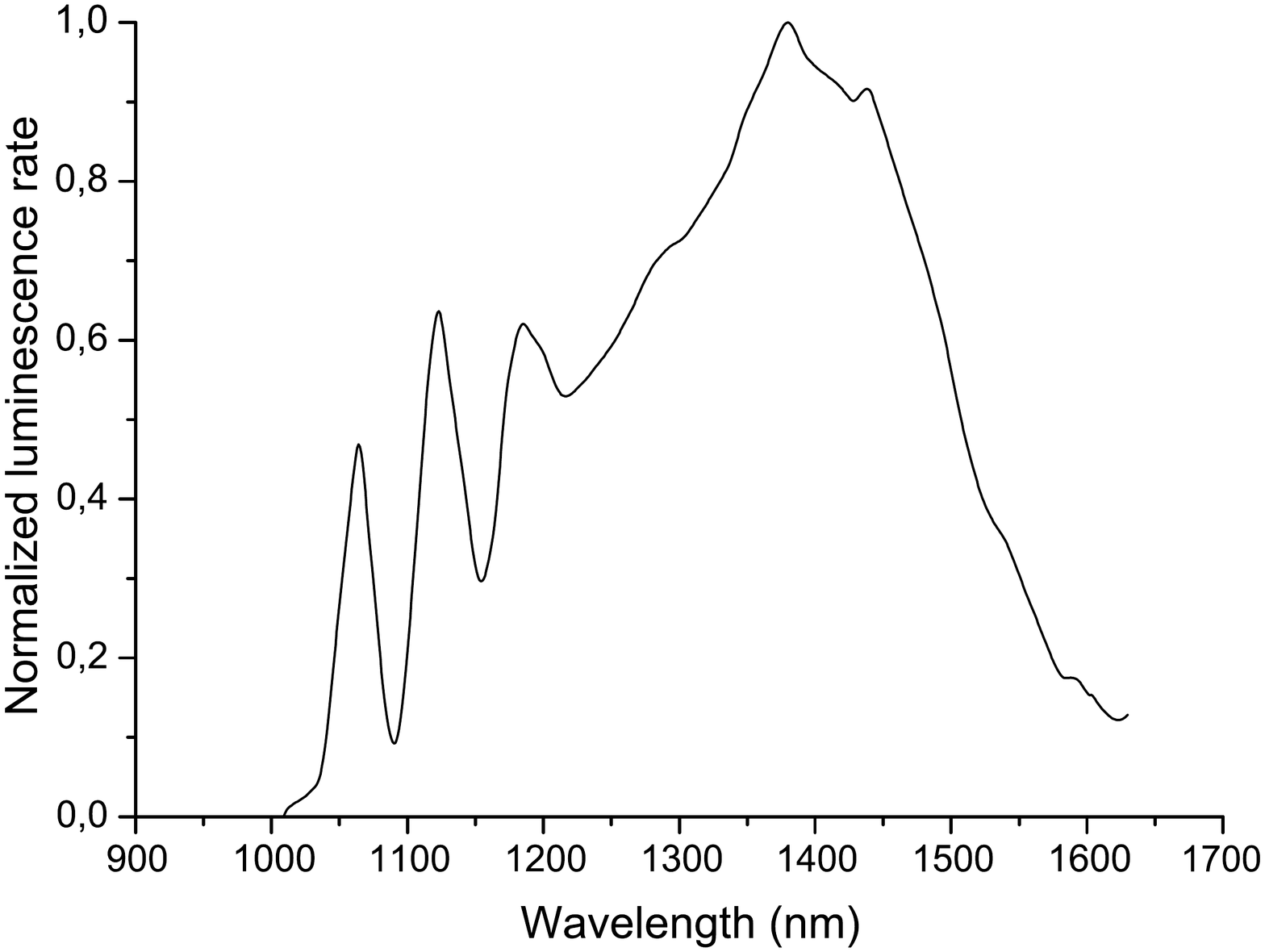}}
\caption{Normalized luminescence spectrum under the excitation  at 532~nm  in the regular GeO2 fiber sample.}
\label{fig:Normalized_luminescence_rate}
\end{figure}

Fig.~\ref{fig:Normalized_luminescence_rate}
illustrates the measured luminescence spectrum in the IR-region. It overlaps completely with expected nongenerated-triplet generation region and covers the whole spectral range of InGaAs-detectors. We assume that peaks at 1000 - 1200~nm are artifacts related to chromatic aberrations and coatings of lenses.


\begin{figure}[t]
\centering
\fbox{\includegraphics
[scale=0.29]{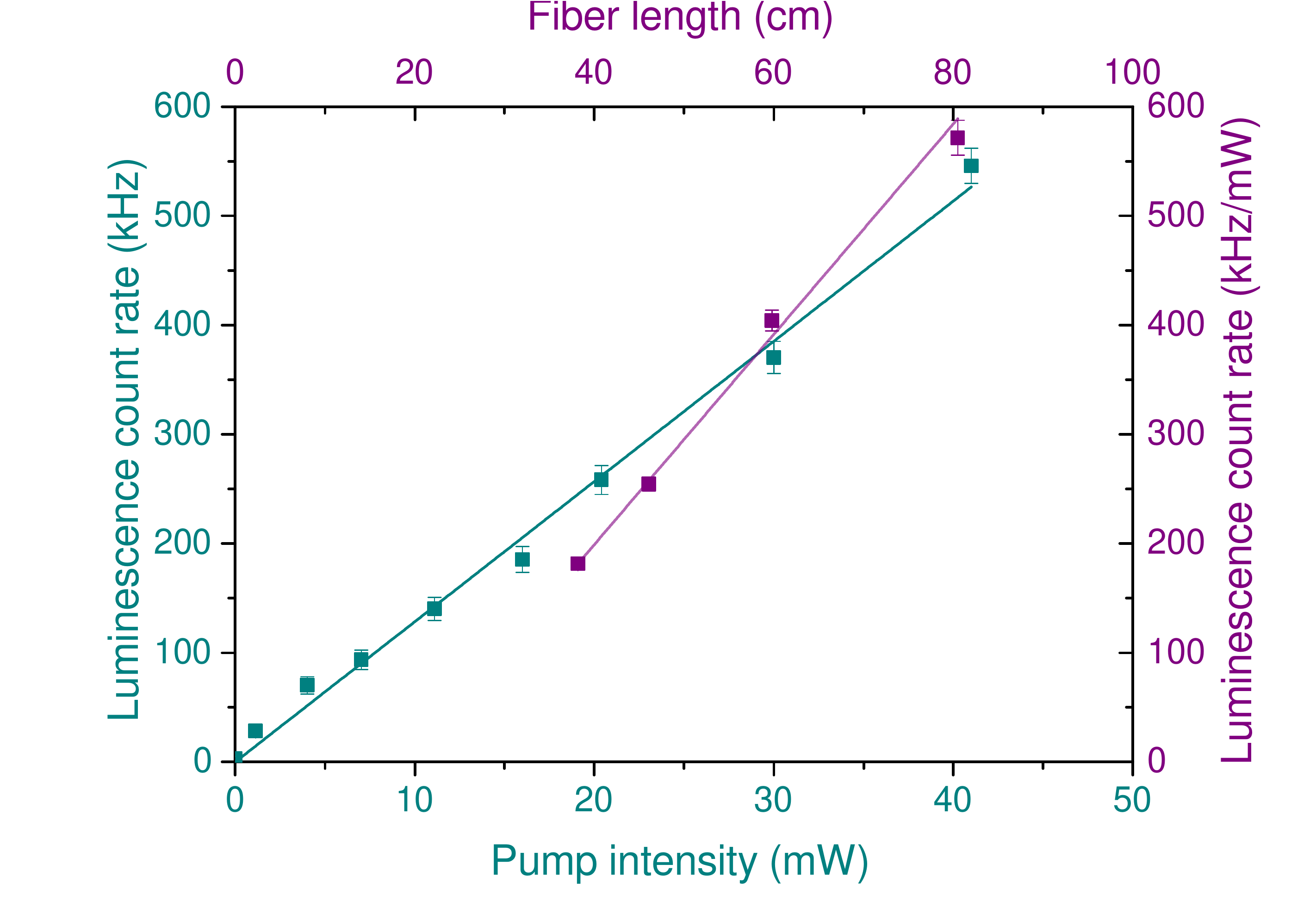}}
\caption{Cyan: the dependency of luminescence count rate on the pump power (dots) and its linear fit (solid line).
Purple: the dependency of luminescence count rate on the fiber length (dots) and its linear fit (solid line).}
\label{fig:lum power vs fiber length & pump int}
\end{figure}

Luminescence intensity depends linearly both on  the pump power
and fiber length (see Fig.~\ref{fig:lum power vs fiber length & pump int}) in the spectral region of interest in context of threephoton generation experiment. This fact actually is in favor of the absence of contribution of nonlinear optical effects.

The integral count rate in the region 1000-1800~nm is $\rm 10^9\ Hz\, W^{-1}\, m^{-1}$ which corresponds to $\rm 10^{11} \ Hz\, W^{-1}\, m^{-1}$ of photon generation rate. This value exceeds the expected triplet generation rate in 12 orders \cite{Richard2011}. Estimations of the time required for reliable triplet registration accompanied with parasitic radiation, similar to \cite{Borshchevskaya2015}, reveal that limitation of triplet registration time to several hours leads to the maximal luminescence rate of $\rm 10^4 \ Hz\, W^{-1}\, m^{-1}$. Consequently, it should be decreased by 7 orders at least.

In order to understand the real location of luminescence source in the fiber (core, cladding or polymeric coating) the coating was removed and the fiber was covered with the immersion fluid (seed oil), possessing the refractive index to be close to the cladding one. Each step allows us to reduce parasitic signal, but not significantly (up to 2 times totally). So we conclude that the significant part of the luminescence origins is located in the fiber core.
%

\begin{figure}[]
\centering
\fbox{\includegraphics
[scale=0.50]
{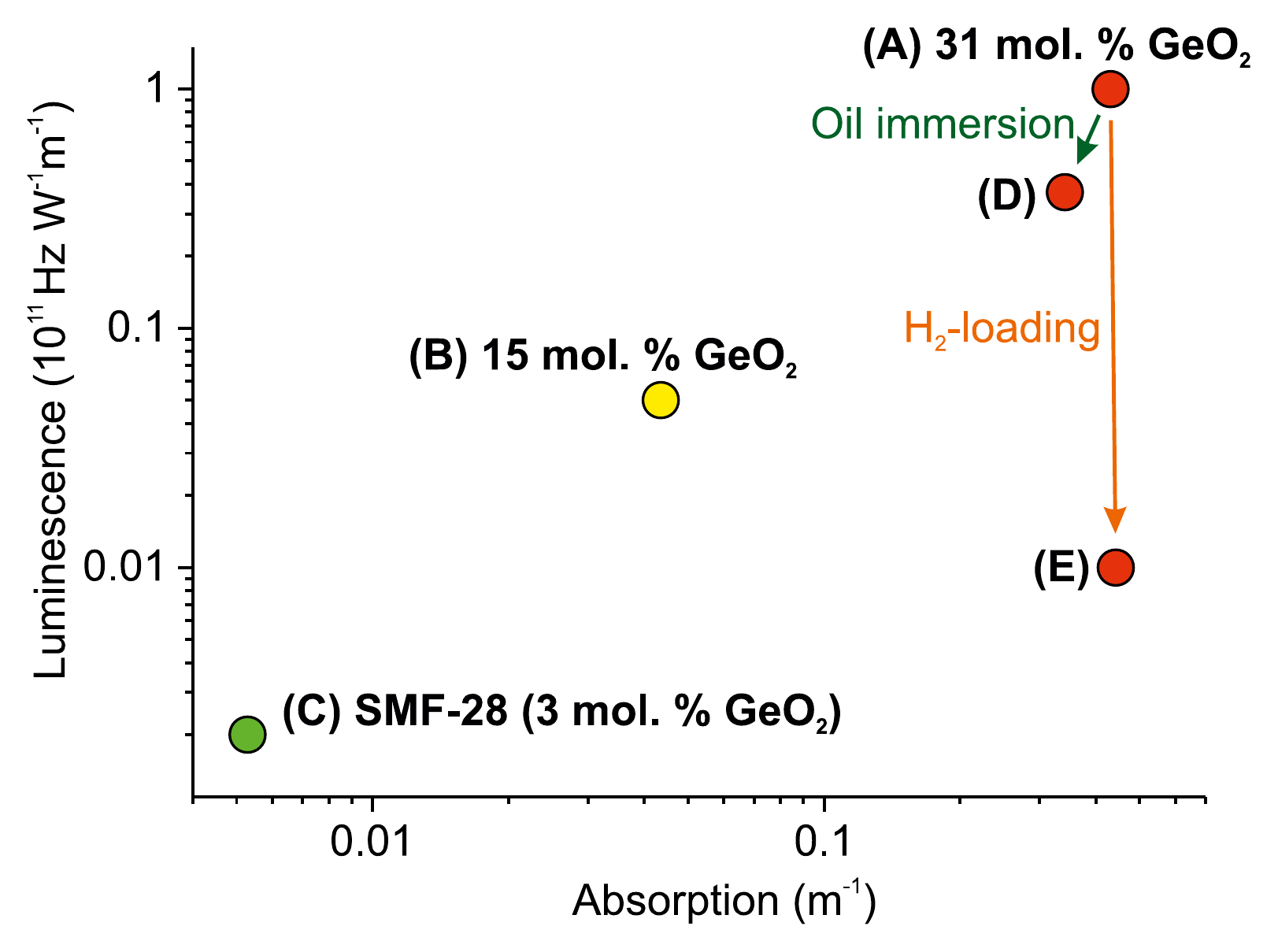}}
\caption{The correlation between absorption and luminescence in fibers with different germanium concentration.}
\label{fig:5}
\end{figure}

To understand the role of $\rm GeO_2$ in the parasitic radiation we tested fiber samples with less germanium concentration and obtained smaller level of luminescence as well as smaller absorption level under the excitation at 532~nm. The obtained absorption and luminescence levels are shown in Fig.~\ref{fig:5}.
In spite of a fiber with optimal for three-photon generation germanium concentration 31~mol.\%$
\rm GeO_2$ (A) we analyzed a fiber with 15~mol.\%
$\rm GeO_2$ (B), a standard SMF-28 with 3~mol.\%
$\rm GeO_2$ (C), a fiber (A) without polimeric coating located inside the immersion fluid (D) and a fiber (A) after the hydrogen saturation (E) which is discussed below.

For a fiber without germanuim we registered neither luminescence (higher than the dark count rate $2\cdot10^4$ Hz)
no absorption at the wavelength 532~nm.

At the same time there was no luminescence in visible and near IR from 400 to 1000~nm.

We notice that shifting the pump wavelength toward the red region leds to the reduction of the luminescence level. For instance, in case of the pump at 650~nm the luminescence intensity decreased in two orders in comparison with the case of the pump at 532~nm.

The influence of bending radius of a fiber on the luminescence level was examined as well. It was shown that the decreasing of bending radius in 4 times leads to the increasing of luminescence signal in $1.7$ times which may be explained by the formation of additional defects during the deformation.

When carrying experiments on registration of luminescence spectra the quality of radiation coupling appeared to be one of the important parameter. It turns out that the pump focusing into fiber in non-optimal way allowed to shift the maximum of luminescence intensity toward the longer wavelengths. However, the complete elimination of the luminescence was not achieved with this technique.

The only effective approach leading to the luminescence suppression is hydrogen loading of fibers under high pressure \cite{Rybaltovskii2008}. Interaction with hydrogen was carried out for two fiber samples, drawn from one preform, under the following conditions:
\begin{itemize}
\item[\--] temperature $20^\circ$C, pressure 100 atm., loading time 24 hours, fiber diameter 123~$\mu$m;
\item[\--] temperature $100^\circ$C, pressure 200 atm., loading time 24 hours, fiber diameter 140~$\mu$m.
\end{itemize}

In both cases the luminescence level decreased in two orders of magnitude and was equal to $\rm 10^7 \ Hz\, W^{-1}\, m^{-1}$. However it still was too high for threephoton registration. It is worthy to notice that after performing hydrogen saturation procedure the intensity of luminescence reduced but the absorption level didn't change, while in all other non-loaded samples the levels of luminescence and absorption behaved identically to each other.

\section{Discussion}

Germania-silica fibers draw the great interest due to the possibility of its application for writing of bragg gratings \cite{Hill1978}, widely used in manufacturing of fiber lasers \cite{Ball1991}, different sensors  \cite{Yun-Jiang1997} as well as telecommunication \cite{Poumellec1996} and dispersion compensation \cite{Williams1994}. Therefore there exist a lot of publications related to the  analysis of absorption and luminescence in these fibers. However, we were unable to find investigations of luminescence in the region 1-2~$\mu$m , excited at 532~nm.

It was demonstrated that a process of glass net saturation with molecular hydrogen under the pressure from 1 to 100 atm. increase photosensitivity of fibers \cite{Meltz1991}. If it is carried out at temperature higher $250^\circ$C than the number of germanium oxygen-deficient centres (GODC) grows \cite{Neustruev1994}. Chemical reactions appear only at defect sites \cite{Stone1987} and in context of current work are caused only by germanium  because in fibers, consisting of pure silica and fluorine only, there was registered no changes in transmission after hydrogen saturation even at high temperatures \cite{Wehr1985}.

Chemical reactions under hydrogen saturation result in additional absorption lines in visible and IR-regions \cite{Stone1987} at the level of tens dB/km \cite{Uesugi1985} due to the formation of SiOH and GeOH. However, if the phosphorus concentration in fibers is low  these losses become  negligible \cite{Stone1987}. Our samples contain no phosphorus, and its length was no more than 10~m, therefore mentioned losses were not remarkable at the presence of luminescence. Besides, it's considered that chemical interaction of hydrogen with glass net is significant only at temperatures higher $250^\circ$C, while at less temperatures (which corresponds to hydrogen saturation conditions applied to samples under current investigation) influence of $\rm H_2$ is equal to a diffusion \cite{Neustruev1994}. A diffusion involves to situations when $\rm H_2$ molecules which didn't chemically react initiate an absorption line at 1.9~$\mu$m \cite{Uesugi1985}, which is out of InGaAs-detectors sensitivity region.

D.~M.~Krol et al. \cite{Krol1991} obtained the luminescence at 720~nm under excitation at 532~nm, but data were received only for wavelength less than 900~nm. The dependence of luminescence intensity on pump intensity was linear similar to our results. The authors assumed these luminescence peak to be connected with GeH-centers. However, hydrogen is nessesary for the existence of these centers while in our case the luminescence is obtained without $\rm H_2$-loading. At the same time we got the opposite dependence of luminescence intensity on the hydrogen saturation degree: in the work \cite{Krol1991} under more strict parameters of $\rm H_2$-loading the luminescence level increased (which is connected with the grows of the number of GeH-centers), while in our experiments, on the contrary, $\rm H_2$-loading leads to the reduction of the luminescence intensity.

H.~Kuswanto et al. \cite{Kuswanto1999} obtained the luminescence from GeH-centers at 740~nm under the excitation at 488~nm. They got that similar to \cite{Krol1991} luminescence level increased with temperature increase, but the difference was that in \cite{Kuswanto1999} at the first stage there was applied $\rm H_2$-loading at the room temperature, than heating while in \cite{Krol1991} temperature was increased simultaneously with the hydrogen absorption. $\rm H_2$-loading at the room temperature without the consequent heating didn't result in the appearance of luminescence  while in our work there existed an influence of $\rm H_2$-loading performed both at the room temperature and at $250^\circ$C.

We notice that the germanium concentration in fiber samples from \cite{Krol1991} was relatively small (2.5~mol.\%
$\rm GeO_2$) while our samples were highly germanium-doped (31~mol.\%
$\rm GeO_2$). But according to \cite{Mashinsky1994} the increase of $\rm GeO_2$ concentration doesn't lead to the appearance of new types of defects in fibers.

One of the well-known absorption peaks in germania-silica fibers is that with maximum at 240~nm (5.12 eV) \cite{Neustruev1994}, which demonstrates the presence of GODC \cite{Dianov1997}. UV-excitation of GODC at 240~nm leads to the appearance of induced losses in UV- and visible region and excitation of paramagnetic centers Ge(n) \cite{Poyntz-Wright1988}. Besides, these centers are related to luminescence lines at 620~nm (2~eV), 680~nm (1.8~eV) and 400~nm (3.13~eV), registrated under two-photon absorption at 488~nm \cite{Kohketsu1989}.

At the same time it was shown that two-photon absorption in germania-silica fibers is possible for photons with wavelengths not longer than 550~nm which corresponds to the one-photon absorption line with maximum at 240~nm \cite{Kuo1990}. The radiation with the wavelength 266~nm corresponding to two-photon absorption at the wavelength of the pump, used in our experiments, 532~nm, also excite partially this line \cite{Hand1990}. Thus, one could assume that the luminescence in our samples is directly caused by two-photon absorption at this line in GODC. However this statement contradicts to the fact of linear dependency of luminescence intensity on the pump power (Fig.~\ref{fig:lum power vs fiber length & pump int}).

It is also known that after the break of regular bonds in silica glass, in spite of oxygen-deficient centers, oxygen-redundant centers appear as well. One of them is non-bridging oxygen (NBO) center which absorbs at 260~nm (4.75~eV). However we were managed to find only the data on the corresponding luminescence at the wavelength 650~nm \cite{Bakos2002}.

In the paper \cite{Poumellec1998} there was observed the luminescence at 270~nm, however, it turned out that it wasn't related to the germanium doping.

\section{Conclusion}

In this Letter we have analyzed the origins of the luminescence in germania-silica fibers with high germanium concentration (31~mol.\%
$\rm GeO_2$) in the region 1-2~$\mu$m with a laser pump at the wavelength 532~nm. It is shown that such fibers demonstrate the high level of luminescence which  unlikely allows the registration of TOSPDC process in such fibers. The only efficient approach to the luminescence reduction -- in 2 orders -- was hydrogen saturation of fiber samples, however, even in this case the level of residual luminescence was still much higher than the threshold  for three-photon detection. Thus, it seems that germania-silica fibers can't be exploited for spontaneous three-photon registration and instead one should use fibers without germanium.

\section{Funding Information}
The work is supported by the Russian Science Foundation under Grant No. 14-02-01338$\Pi$.






%
%

\bibliographystyle{plain}
\bibliography{sample}




\end{document}